# Development of high critical current density in multifilamentary round-wire $Bi_2Sr_2CaCu_2O_{8+\delta}$ by strong overdoping[1]

T. Shen, J. Jiang, A. Yamamoto, U. P. Trociewitz, J. Schwartz, E.E. Hellstrom, and D.C. Larbalestier

*Applied Superconductivity Center, National High Magnetic Field Laboratory, Florida State University, Tallahassee, FL, 32310, USA*

**Abstract:**

$Bi_2Sr_2CaCu_2O_{8+\delta}$ is the only cuprate superconductor that can be made into a round-wire conductor form with a high enough critical current density $J_c$ for applications. Here we show that the $J_c$(5 T,4.2 K) of such Ag-sheathed filamentary wires can be doubled to more than $1.4 \times 10^5$ A/cm$^2$ by low temperature oxygenation. Careful analysis shows that the improved performance is associated with a 12 K reduction in transition temperature $T_c$ to 80 K and a significant enhancement in intergranular connectivity. In spite of the macroscopically untextured nature of the wire, overdoping is highly effective in producing high $J_c$ values.

---

Perhaps the most pressing wish of the large scale superconducting applications community is to develop a round wire multifilamentary conductor from a 100 K class superconductor, which could viably replace the two Nb-based materials, Nb-Ti and Nb$_3$Sn with $T_c$ of 9 and 18 K, from which virtually all superconducting applications are presently made.[1] For more than 20 years the dream of using cuprate high temperature superconductors (HTS) with $T_c$ in the 90 to 110 K range has fueled a conviction that helium-free magnets operating at much higher temperatures are possible. The three available HTS conductors all offer the possibility to generate magnetic fields far beyond the maximum of ~22 T possible with Nb$_3$Sn, since cuprates have critical fields at 4.2 K greater than 100 T, even in the inferior direction. But the major obstacle to their use is the tendency of cuprate grain boundaries (GBs) with misorientation angle $\theta$>3-4º to have depressed $J_c$ due to local GB suppression of the carrier density and the superconducting order parameter.[2, 3] Thus (Bi,Pb)$_2$Sr$_2$Ca$_2$Cu$_3$O$_x$ (Bi-2223) conductors have never achieved their full potential because only a partial uniaxial texture with FWHM ~10-12º can be developed,[4] while coated conductors of YBa$_2$Cu$_3$O$_{7-\delta}$ (YBCO) show much higher $J_c$ because a strong biaxial texture with FWHM<5º can be developed by epitaxial or seeded growth.[5-7] But large aspect ratio tapes, ~20:1 for Bi-2223 or typically 4000:1 for YBCO, are far from optimum, because it is hard to cable flat tape conductors and tapes have large hysteretic losses in perpendicular magnetic field **H**. Since the cuprate GB problem is widely believed to be intrinsic to their small carrier density and proximity to a parent, antiferromagnetic insulating state, understanding the remarkable properties of round-wire, Ag-sheathed Bi-2212 conductor has quite general importance. Unlike any other cuprate, Bi-2212 can attain high $J_c$ in round wires which lack long-range texture. This letter addresses the final process step that greatly enhances their $J_c$ and ties it to oxygenation treatments that overdope the Bi-2212 phase in ways that are generally not possible in Bi-2223 or YBCO.

Bi-2212 round wire is composed of a myriad of ~100-200 μm long and ~0.1-0.3 μm thick plate-like Bi-2212 grains, often arranged in 1-5 μm thick colonies that share a common $c$-axis with [001] twist boundaries[8]. Although the $c$-axis is often aligned perpendicular to the wire axis, there is no azimuthal texture. Despite this absence of long-range texture, powder-in-tube (PIT) Bi-2212 round wire can carry remarkably high $J_c$ values (~$1 \times 10^5$ A/cm$^2$ at 45 T and 4.2 K)[9]. The combination of high $J_c$ and poor texture suggests that GB transport in Bi-2212 round wire is much easier than in Bi-2223 and YBCO.

Partial-melt processing (see the typical multistep process in Fig. 1) is vital to develop high $J_c$.[10] The conductor whose filament powder starts as essentially single-phase Bi-2212 is heated above the Bi-2212 peritectic temperature to melt the filaments, producing a liquid containing all elements including Ag, alkaline earth cuprate (($Sr,Ca)_{14}Cu_{24}O_x$), and copper-free phase ($Bi_9(Sr,Ca)_{16}O_x$) mixture. Slow cooling, in this case 2.5 °C/hr, solidifies this mixture and below ~872 °C Bi-2212 grains nucleate. Much discussion on improving conductor $J_c$ deals with manipulating the melt phase assemblages.[11] Here, we concentrate on the final portion of the heat treatment, which occurs after forming the Bi-2212 network, where we enhance the superconducting connectivity by slow cooling in an $O_2$-rich atmosphere.

We quenched 4 cm long sections of a Ag-sheathed wire containing 7 bundles of 85 Bi-2212 filaments fabricated by Oxford Superconducting Technology[9,12] at multiple points (Q836C means quenching from 836 °C) in the process using brine as the quench medium to preserve the high temperature microstructures and electromagnetic properties, without introducing damage that might reduce $J_c$.[13] Thus we could directly correlate the superconducting properties to the high temperature state.

Microstructures were carefully examined and phase chemistry was determined using a field emission scanning electron microscope. The important point is that no observable change in the phase state occurred below the highest temperature examined here, 836 °C. $T_c$ was evaluated from zero-field-cooled magnetic moments measured in a SQUID magnetometer on 5 mm long samples with the wire axis parallel to **H**. The irreversibility field $H_{irr}$ was approximated by linear extrapolation of the Kramer function $\Delta M^{0.5}H^{0.25}$ to zero, defining $H_K$. $\Delta M$ is the hysteretic magnetization, which is proportional to $J_c$. 5 mm long samples were measured in a 14 T vibrating sample magnetometer with the wire axis perpendicular to **H** so that currents propagate along the wire axis across many GBs. The inter- and intra-grain contributions to the hysteretic moment $\Delta M$ were deduced from the remanent moment $m_R(H_a)$, determined in the SQUID magnetometer by exposing sample to incrementally increasing magnetic field $\mathbf{H}_a$ followed by removal of the field and measurement of the remanent moment $m_R$. Magnetic flux first enters at weak regions such as GBs and finally into the grains, $m_R(H_a)$ in each case being given by the product of the screening currents $I_c$ and the length scale of these currents.[14, 15] Differentiation of $m_R(H_a)$ often shows two distinct peaks corresponding at low fields to *intergrain* currents circulating across GBs, while the higher-field peak corresponds to a combination of *intragrain* currents of high $J_c$ and/or well connected current paths with long length scales. The transport $J_c$ was determined at an electric field criterion of $10^{-6}$ V/cm with field perpendicular to the wire axis at 4.2 K using the Bi-2212 cross-section before reaction as the normalizing area.

Figure 1 compares the transport $J_c$ at self field and 5 T at 4.2 K for each of the 3 quenched samples (836, 650, and 330 °C), together with a fully processed sample (FP) and the sample that was quenched from 836 °C and then given a final low temperature post-anneal (400 °C, 30 hr) in 1 bar flowing $O_2$ (Q836C+PA). We emphasize that there was no visible difference in the phase state and grain structure of these samples, since the Bi-2212 conversion process was complete at 855 °C, before any quenching. Fig. 1 shows that this slow cool at 170 °C/hr in 1 bar flowing $O_2$ from 836 °C dramatically enhanced $J_c$ (4.2 K, 5 T) from 0.7 to $1.4 \times 10^5$ A/cm$^2$. Self field and 5 T $J_c$ are raised similarly.

Figure 2 shows that as the quench temperature decreases, the transition loses its onset kink and $T_c$ monotonically decreases from ~92 K (836 °C) to ~83 K (< 480 °C), indicating strong oxygen pickup, since $T_c$ of cuprates is a parabolic function of hole concentration that increases with oxygen content δ in $Bi_2Sr_2CaCu_2O_{8+\delta}$.[16] δ strongly depends on temperature, increasing from 0.2 at 830 °C to 0.25 at 300 °C in 1 bar oxygen.[16] However, all transitions are broad, partly due to small filament dimensions (~20 µm), as well perhaps due to residual compositional inhomogeneities, preferential flux penetration at high-angle grain boundaries, $Bi_2(Sr,Ca)_2CuO_x$ (Bi-2201) intergrowths, or other secondary phases, and voids. The Q836C+PA sample has the lowest $T_c$ with an onset of 80 K, indicating that it has the highest oxygen concentration.

Figure 3 plots $H_K(T)$ which exhibits the usual behavior of $H_{irr}(T)$, increasing steeply around 20 K.[17] At 20 K, $H_{irr}(T)$ increased from ~5.6 T for Q836C to 7.4 T for fully processed wire, while the Q836C+PA sample shows the highest $H_K(20\ K)$ of 8.1 T. This enhancement is explained by a reduced intra-grain electronic anisotropy brought on by the increased carrier density of the overdoped state.[18, 19]

Figure 4 shows that the remanent current flow paths produce two well separated peaks in $dm_R/d\log H_a$. It is striking that low temperature oxygenation preferentially enhances the first peak in which intergrain paths dominate. The clear implication is that the long-range current flow across GBs is enhanced by oxygenation.

Our central result is that low temperature oxygenation of a macroscopically untextured, round-wire multifilamentary Bi-2212 conductor produces a more than 2 fold enhancement of the in-field $J_c$ (Fig. 1). These treatments enable the high $J_c$ values needed for very-high-field magnets, as demonstrated by the generation of 2.5 T in a 20 T background field[20] and 1 T in a 31 T background field[21], 32 T being a field more than 50% higher than can be generated with any Nb-based magnet. This oxygen pick up overdopes the Bi-2212 phase, decreasing $T_c$ from 92 to 80 K (Fig. 2), but in all other respects enhancing the superconducting properties (Figs. 1-4). Especially valuable to $J_c$ may be the connectivity enhancement shown in Fig. 4.

The mechanisms of decreased current transport through planar cuprate GBs have been extensively studied.[2, 3, 5] Cuprate GBs develop an increasingly suppressed superconducting order parameter (OP) and $J_c$ as $\theta$ increases.[3] This OP suppression is amplified by the extra ionic charge, band bending, and strain-driven $O_2$ depletion in the vicinity of the GB, all of which lead to the GB being underdoped with respect to the grains and closer to the parent non-superconducting state. Neither Bi-2223 nor YBCO can be more than lightly overdoped, thus their GBs are underdoped. The benefits of overdoped GBs are seen in the properties of Ca-doped YBCO, because Ca does allow carrier overdoping of the GB[22-24], and also in some bulk Bi-2212 bicrystal studies[25]. The striking present result is that overdoping an untextured Bi-2212 wire makes a hugely positive influence on $J_c$. The poorly oxygenated Q836C and Q650C wires show a characteristic shoulder at 60-82 K in the $m(T)$ curves (Fig. 2), suggesting a decreased $T_c$ at hole-deficient Bi-2212 grain boundaries. Lower-temperature oxygenation removes this shoulder and sharpens the $T_c$ transition, while reducing the $T_c$ onset. We emphasize that the resulting $J_c$ values of $10^5$ A/cm$^2$ or more are practical values that now enable the next generation of very high field magnets,[20, 21] which makes this overdoping route to a round wire conductor of great practical and scientific interest.

This work was supported by U. S. National Science Foundation Division of Material Research through DMR-0654118 and the State of Florida. The authors are grateful to Van S. Griffin, Natanette C. Craig, and Bill Starch for technical assistance.

Figure Captions:

Fig.1 (Color online) Significant increases occur in the self field and 5 T, 4.2 K transport $J_c$ of Bi-2212 round wire in the final stage of the partial-melt process (inset). In flowing 1 bar $O_2$, samples were melted at a maximum temperature of 894 ºC and slowly cooled at 2.5 ºC/hr to 836 ºC, where they were annealed for 48 hr before being cooled at 170 ºC/hr to room temperature. Samples were quenched at 836 ºC, 650 ºC, and 330 ºC. A fully-processed (FP) wire and an oxygen-rich sample, which was quenched at 836 ºC then post annealed at 400 ºC for 30 hr in 1 bar flowing $O_2$, are also shown. These samples are referred to Q836C, Q650C, Q330C, FP, and Q836C+PA. The dashed lines are given to guide the eye.

Figure 2 (Color online) Zero-field-cooled magnetic moments of Ag-sheathed Bi-2212 multifilament round wire Q836C, Q650C, Q330C, FP, and Q836C+PA induced by warming in a field of 1 mT applied parallel to the wire axis, indicates that significant oxygen overdoping occurs during the final cooling to room temperature.

Figure 3 (Color online) Kramer irreversibility field as a function of temperature for Q836C, Q650C, Q330C, FP, and Q836C+PA. Inset shows our method of determining $H_K$.

Figure 4 (Color online) Dependence of the remanent magnetic moment $m_R$ (inset) and its derivative for the Bi-2212 round wires as measured for increasing fields applied parallel to the wire axis at 5 K. Note that the greatest connectivity is shown by sample Q836C+PA.

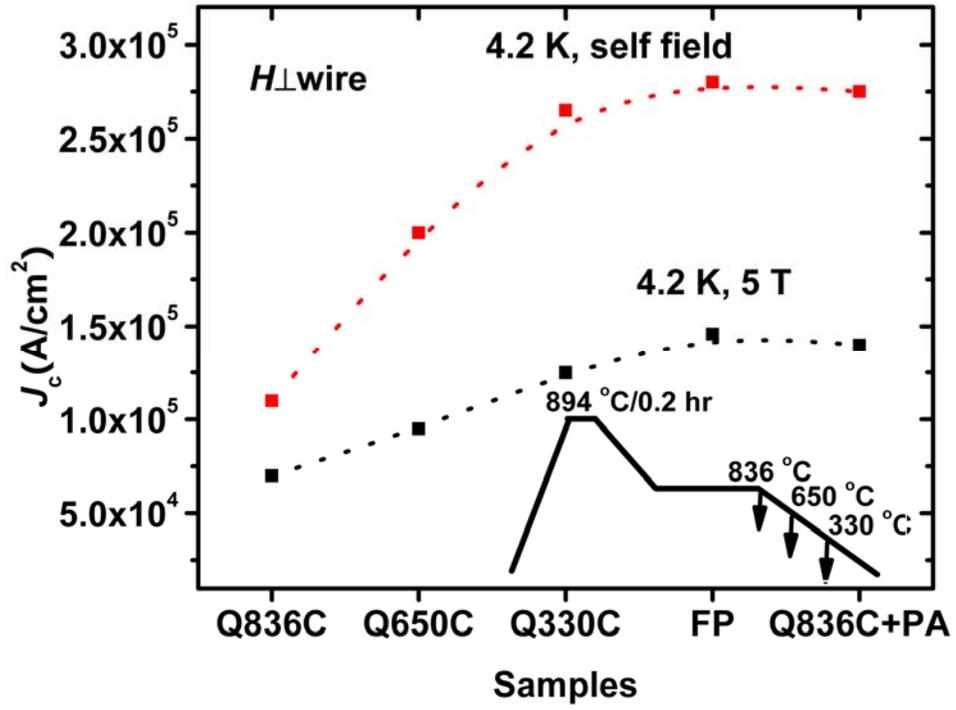

**Figure 1**

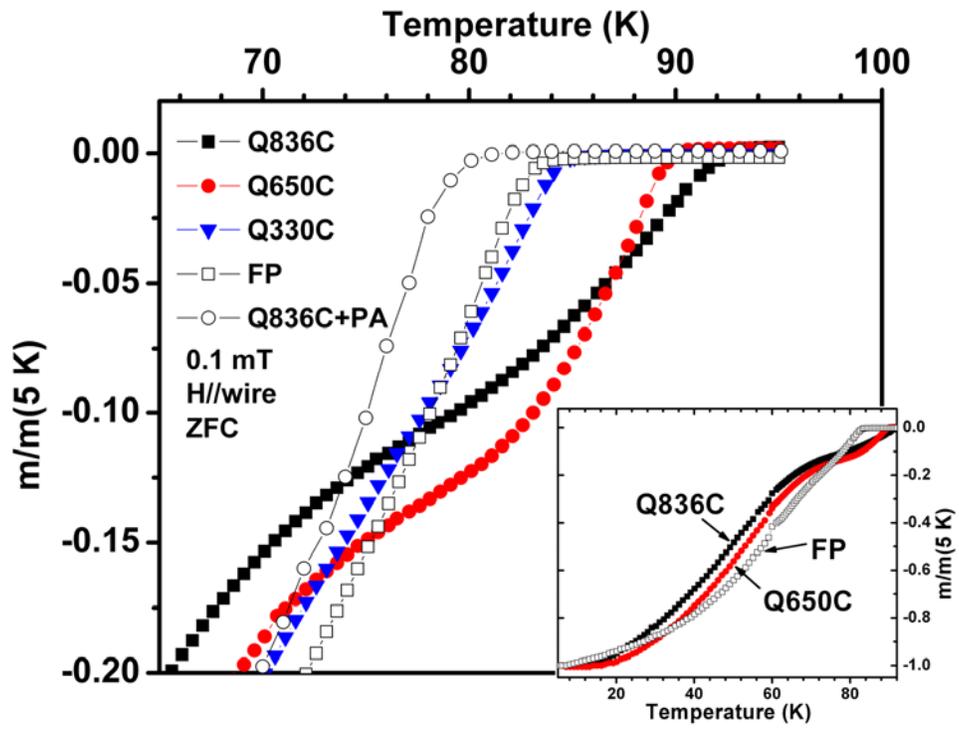

**Figure 2**

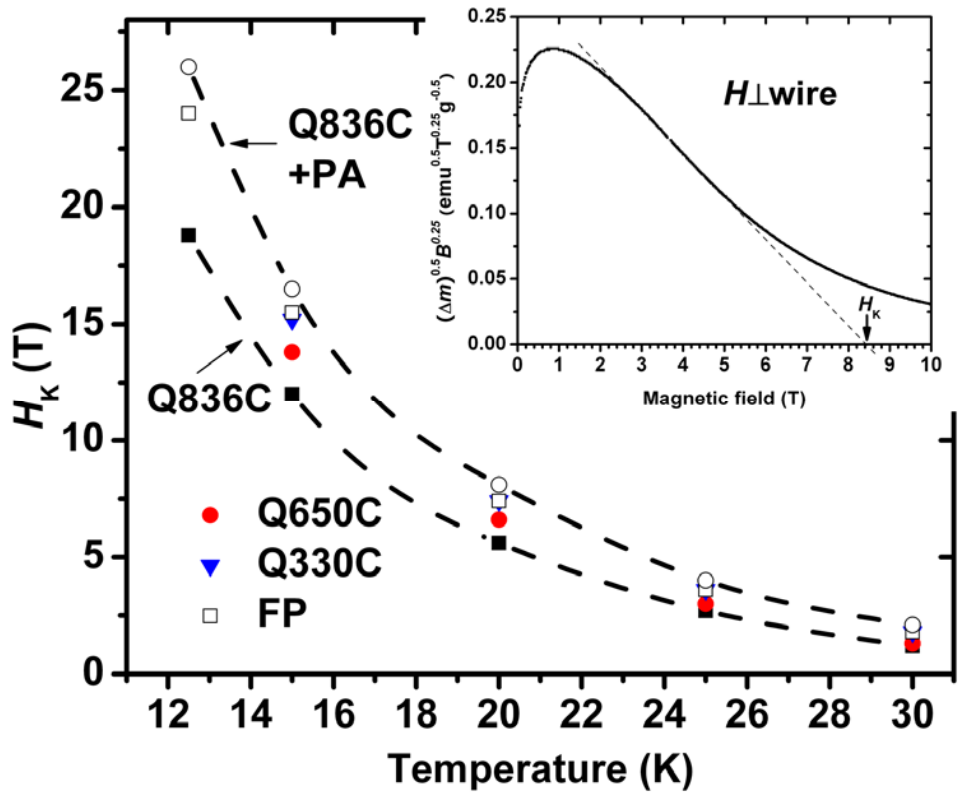

**Figure 3**

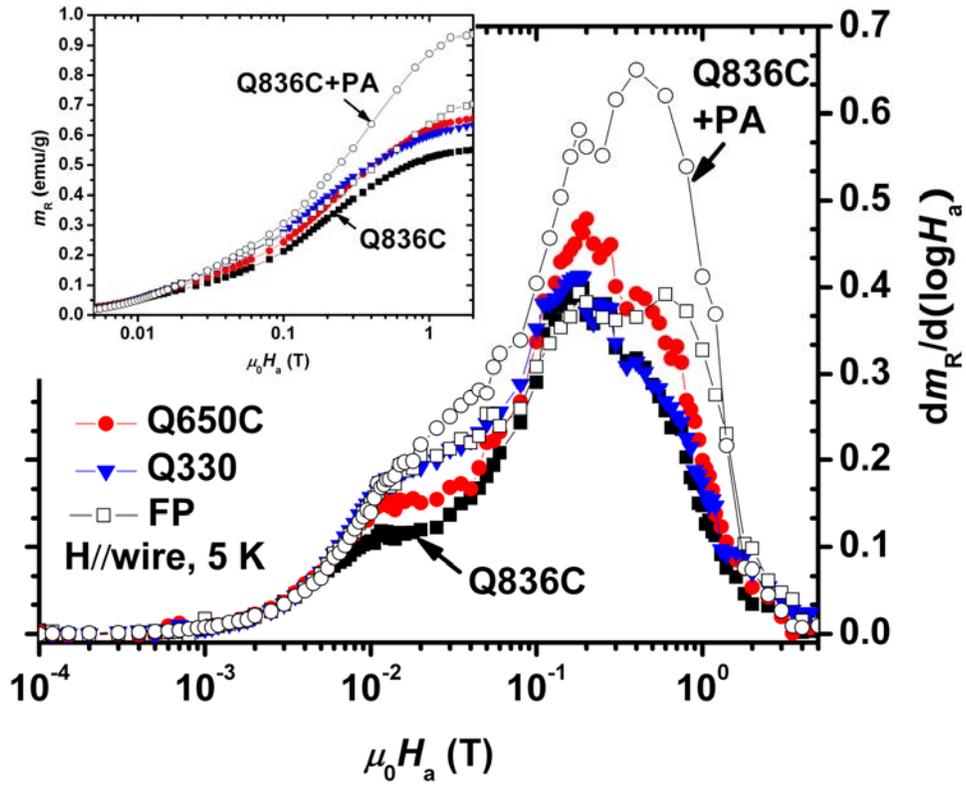

**Figure 4**